\documentclass[aps,tightenlines,nofootinbib,prd]{revtex4}
\pdfoutput=1
\usepackage{amsmath,amscd,amssymb}
\usepackage{color,epsfig}
\usepackage{xfrac}
\usepackage{latexsym}
\usepackage{mathrsfs}
\usepackage{graphicx}
\usepackage{bbm}      % double-striked fonts for sets of integer numbers etc.
\newcommand{\imu}{{\rm i}}

\begin{document}

\title{Quantum Corrections to Solitons in the $\mathbf{\phi^8}$ Model}

\author{I. Takyi$^{a)}$, M. K. Matfunjwa$^{b,c)}$, H. Weigel$^{d)}$}
\affiliation{$^{a)}$Mathematics Department, Kwame Nkrumah University of Science and 
Technology, Private Mail Bag, University Post Office, KNUST-Kumasi, Ghana\\
$^{b)}$African Institute for Mathematical Sciences, 6 Melrose Road, Muizenberg, 7945 Cape Town, South Africa\\
$^{c)}$ Department of Physics, University of Eswatini, Private Bag 4, Kwaluseni M201, Eswatini\\
$^{d)}$Institute for Theoretical Physics, Physics Department, 
Stellenbosch University, Matieland 7602, South Africa}

\begin{abstract}
We compute the vacuum polarization energy of kink solitons in the $\phi^{8}$
model in one space and one time dimensions. There are three possible field potentials
that have eight powers of $\phi$ and that possess kink solitons. For these different 
field potentials we investigate whether the vacuum polarization destabilizes the 
solitons. This may particularly be the case for those potentials that have degenerate 
ground states with different curvatures in field space yielding different thresholds for 
the quantum fluctuations about the solitons at negative and positive spatial infinity. 
We find that destabilization occurs in some cases, but this is not purely a matter of
the field potential but also depends on the realized soliton solution for that potential.
One of the possible field potentials has solitons with different topological charges. 
In that case the classical mass approximately scales like the topological charge. Even 
though destabilization precludes robust statements, there are indications that the vacuum 
polarization energy does not scale as the topological charge. 
\end{abstract}
%\pacs{03.65.Ge, 05.45.Yv, 11.10.Lm}

\maketitle

\section{Introduction}
Solitons are stable, localized particle-like solutions to non-linear wave equations in field 
theories~\cite{Ra82,Manton:2004tk,Vachaspati:2006zz}. These particle-like solutions include 
solitons such as monopoles in $(3+1)$ dimensions and vortices, strings and lumps in $(2+1)$ 
dimensions.  The soliton solutions have wide applications in many branches of physics: in 
cosmology~\cite{Vilenkin:2000jqa}, condensed matter physics~\cite{Schollwock:2004aa,Nagasoa:2013},
as well as hadron and nuclear physics~\cite{Weigel:2008zz}. The interested reader is
directed to those textbooks and review articles for more details and further references.

In one space and one time dimensional models with distinct vacua, soliton solutions connect
neighboring vacua between negative and positive spatial infinity. As there is no finite
energy continuous transformation between these vacua at spatial infinity, these so-called 
kink solitons are topologically stable. When embedded in higher dimensions, these kinks 
emerge as branes or interfaces. In one space dimension dynamical stability can be argued 
for from Derrick's theorem~\cite{De64}.  The derivative and potential contributions to 
the classical energy scale oppositely when varying the extension of the kink. The derivative 
part diverges when the static configuration is shrunk to a point, so does the potential part 
for wide configurations. Hence the classical energy is minimized for a finite size of the 
kink. The situation may, however, change when nonlocal quantum effects are included. Here we will
therefore compute the leading (one-loop) quantum correction to the classical energy of the kink 
using spectral methods~\cite{Graham:2009zz}. Such corrections have also been computed 
within the heat kernel expansion with $\zeta$-function
regularization~\cite{AlonsoIzquierdo:2011dy,AlonsoIzquierdo:2012tw} for other soliton models.
Agreement is observed whenever comparison is possible. However, spectral methods are 
technically less involved and do not require the truncation of an expansion.

Generally non-linear field theories have degenerate vacua with different curvatures 
in field space. The different curvatures translate into different masses for the fluctuations
about the degenerate vacua. We will call {\it primary} vacuum the one with the lowest of 
all allowed masses, the others will be called {\it secondary} vacua. In models with 
several fields the primary vacuum contains the smallest of all curvature eigenvalues. It has 
recently been conjectured that the occurrence of such secondary vacua causes the solitons 
to be unstable on the quantum level. As the kink occupies the secondary vacuum over 
an increasing region in space, the one-loop quantum energy decreases without a lower bound 
and thereby destabilizes the kink. This conjecture has been drawn from two model calculations, 
in the $\phi^6$ model \cite{Weigel:2016zbs,Weigel:2017iup} and the multi-field Shifman-Voloshin
model~\cite{Weigel:2018jgq,Weigel:2019rhr}. In the context of the $\phi^6$ model a similar 
conclusion was drawn when it was observed that fluctuations produce a net force on the 
kink~\cite{Romanczukiewicz:2017hdu}. 

In this paper we will consider the $\phi^8$ model that also contains topologically stable 
solitons ~\cite{Khare:2014kva,Gani:2015cda} and which, for certain model parameters, exhibits 
secondary vacua. Recently this model has attracted quite some attention, {\it cf.} 
Ref.~\cite{Christov:2020zhb} and references therein, as it induces long range interactions
in the kink-antikink scattering. In this model will find that indeed the vacuum polarization 
energy (VPE), which is the leading quantum correction to the soliton static energy, is not 
bounded from below in certain versions of this model either. We note that though the $\phi^8$
is super-renormalizable (as any scalar theory in one space and one time dimensions) perturbatively, 
we will see that renormalizability is nevertheless an issue and that only a particular renormalization 
condition (the no-tadpole prescription) leads to a finite quantum correction. Super-renormalizability 
makes statements on the number the divergent Feynman diagrams but not the structure of the counterterms.

In the next section we briefly outline the construction principle of soliton solutions in one(space) 
dimensional models and list the field potentials for our present study. In section \ref{sec:vpe} we 
describe the general formalism for computing the one-loop energy correction to the energies of the 
solitons that we construct explicitly in section \ref{sec:solitons}. We present and discuss our 
numerical results in section \ref{sec:results} and conclude in section \ref{sec:conclude}.

\section{The Models}
\label{sec:models}

Starting point is the general Lagrangian for a $D=1+1$ dimensional model
\begin{equation}
\mathcal{L}=\frac{1}{2}\partial_\mu\phi\partial^\mu\phi-U(\phi)\,,
\label{eq:deflag}
\end{equation}
where $U(\phi)$ is the field potential that governs the dynamics.
In one space dimension it is standard to derive a first order differential equation
for the soliton profile \cite{Ra82}. Integrating the right-hand-side in
\begin{equation}
x-x_0=\pm\int_{\phi_K(x_0)}^{\phi_K(x)}\, \frac{d\phi}{\sqrt{2U(\phi)}}
\label{eq:genkink4deg}
\end{equation}
then yields an implicit function for the profile function $\phi_K(x)$. The first order
formalism furthermore simplifies the classical energy (mass) 
\begin{equation}
E_{\rm cl} =M = \int_{-\infty}^\infty dx\, \left[
\frac{1}{2}\left(\frac{d\phi_K}{dx}\right)^2+U(\phi_K)\right]=
\int_{\phi_K(-\infty)}^{\phi_K(\infty)}d\phi\,\sqrt{2U(\phi)}
\label{eq:ecl}
\end{equation}
for the soliton solution. For the $\phi^8$ model we consider three different field
potentials
\begin{equation}
U_2(\phi)=\lambda^2\left(\phi^2-a^2\right)^2\left(\phi^2+b^2\right)^2\,,\quad
U_3(\phi)=\lambda^2\left(\phi^2-a^2\right)^2\left(\phi^2+b^2\right)\phi^2\quad {\rm and}\quad
U_4(\phi)=\lambda^2\left(\phi^2-a^2\right)^2\left(\phi^2-b^2\right)^2\,.
\label{eq:field1}
\end{equation}
The subscripts denote the number of degenerate vacua that the respective potential contains.

Before discussing the model and its soliton solutions in more detail we will briefly 
review the computation of the VPE, in particular in the context that secondary vacua 
exist.

\section{Vacuum polarization energy}
\label{sec:vpe}

The VPE, $E_{\rm vac}$ emerges from the shift of the zero point energies of the fluctuations in the 
presence of the soliton. This shift manifests itself in two contributions, from the discrete
bound states (b.s.) and from the modified density of the scattering states $\Delta\rho(k)$.
Here $k$ is the wave number of the continuous scattering states. We thus write
\begin{equation}
E_{\rm vac}=\frac{1}{2}\sum_j^{\rm b.s.} E_j 
+\frac{1}{2}\int_0^\infty dk\,\sqrt{k^2+m^2_L} \Delta\rho(k) +E_{\rm ct}\,,
\label{eq:vpe1}
\end{equation}
where $m_L$ is the mass for fluctuations about the primary vacuum and $E_{\rm ct}$ 
is the contribution from the counterterms that render the model finite at one-loop order.
We extract both, the bound state energies $E_j$ and $\Delta\rho(k)$ from the fluctuations 
$\eta(t,x)$ about the soliton. Since the soliton is static we can separate the time 
dependence as $\eta(t,x)={\rm e}^{\imu \omega t}\eta(x)$. We then linearize the field
equations for a general field potential $U(\phi)$.
\begin{equation}
\left[-\frac{d^2}{dx^2} + V(x) \right] \eta(x) = \omega^2 \eta(x).
\qquad {\rm where}\qquad 
V(x) = \left. \frac{d^2U(\phi)}{d\phi^{2}}  \right|_{\phi=\phi_{K}(x)}.
\label{eq:waveequa}
\end{equation}
As described at the end of this section, it
is straightforward to (numerically) obtain the discrete bound state solutions $\omega=E_j$ 
with $0\le E_j\le m_L$. The continuous scattering states are parameterized by real 
$k=\sqrt{\omega^2-m_L^2}$, Unfortunately, extracting scattering data and subsequently $\Delta\rho(k)$ 
is subtle when there are secondary vacua. In such a case the background potential $V(x)$ 
approaches different values at negative and positive spatial infinity. We take our frame 
of reference such that $\lim_{x \rightarrow -\infty}V(x)=m_L^2$ and define 
$m_R^2=\lim_{x \rightarrow +\infty}V(x)$ so that always $m_R\ge m_L$. 

According to the Krein formula~\cite{Faulkner:1977zz} we find 
\begin{equation}
\Delta \rho(k) = \frac{1}{\pi} \frac{d\delta(k)}{dk}
\qquad {\rm where}\qquad 
\delta(k)=-\frac{\imu}{2} {\rm log}\left[{\rm det}S(k)\right]
\label{eq:phase}
\end{equation}
is the sum of the eigenphase shifts that we extract from the scattering
matrix\footnote{Branches of the logarithm must be taken such that $\delta(k)$ is 
a smooth function with $\delta(\infty)=0$.} $S(k)$. To compute $S(k)$ we first define
the pseudo-potential
\begin{equation}
V_{p}(x)=V(x)-m_{L}^{2}+(m_{L}^2+m_{R}^2)\Theta(x-x_{m})
\label{eq:pseudopot}
\end{equation}
where $x_m$ is an arbitrary matching point. Then the wave-equation reads
\begin{equation}
\Big[-\frac{d^2}{dx^2}+V_p(x)\Big] \eta(x) =
\begin{cases}
k^2 \eta(x), & \quad x \leq x_m\\
q^2 \eta(x), & \quad x \geq x_m\,,
\end{cases}
\label{eq:modwvequa}
\end{equation}
with $q=\sqrt{k^2+m_{L}^{2}-m_{R}^{2}}$. For real $q$, {\it i.e.} 
$k\ge\sqrt{m_R^2-m_L^2}$ we formulate a variable phase 
approach~\cite{Cal67} by parameterizing
\begin{align}
& x\leq x_m: \quad \eta(x) = A(x)e^{\imu kx}  \qquad \quad A^{\prime\prime}(x)=
-2\imu kA^{\prime}(x)+V_p(x)A(x) \nonumber \\
& x\geq x_m: \quad \eta(x) = B(x)e^{\imu qx}  \qquad \quad B^{\prime\prime}(x)=
-2\imu qB^{\prime}(x)+V_p(x)B(x)\,.
\label{eq:parametequa} 
\end{align}
Here, and subsequently, a prime denotes the derivative with respect to the coordinate  $x$.
We solve Eq.~(\ref{eq:parametequa}) with boundary conditions 
$A(-\infty)=B(\infty)=1$ and $A^{\prime}(-\infty)=B^{\prime}(\infty)=0$ yielding 
the scattering matrix~\cite{Kiers:1996jt}
\begin{equation}
S(k)=\begin{pmatrix}
{\rm e}^{-iqx_m} & 0 \cr 
0 & {\rm e}^{ikx_m}
\end{pmatrix}
\begin{pmatrix}
B & -A^\ast \cr
iqB+B^\prime & ikA^\ast-A^{\prime\ast}
\end{pmatrix}^{-1}
\begin{pmatrix}
A & -B^\ast \cr
ikA+A^\prime & iqB^\ast-B^{\prime\ast}
\end{pmatrix}
\begin{pmatrix}
{\rm e}^{ikx_m} & 0 \cr 
0 & {\rm e}^{-iqx_m}
\end{pmatrix}\,,
\label{eq:Smatrix}
\end{equation}
where $A=A(x_m)$, etc. are the coefficient functions at the matching point. In 
the second case, $k \le \sqrt{m_{R}^{2}-m_{L}^{2}}$ we replace $\imu q$ by 
$\kappa = \sqrt{m_{R}^{2}-m_{L}^{2}-k^2} \ge 0$ and parameterize the wave function 
for $x\ge x_m$ as $\eta(x) = B(x)e^{-\kappa x}$. In that region the differential 
equation now reads $B^{\prime\prime}(x)=\kappa B^{\prime}(x)+V_p(x)B$ and we extract
the reflection coefficient as
\begin{equation}
S(k) = -\frac{A\left(B^{\prime}/B- \kappa - \imu k\right)-A^{\prime}}
{A^\ast\left(B^{\prime}/B-\kappa + \imu k\right)-A^{\prime\ast}} e^{2 \imu kx_m}\,.
\label{eq:reflecoeff}
\end{equation}
In the non-tadpole renormalization scheme, which is the only one applicable when 
secondary vacua emerge \cite{Weigel:2016zbs}, the counterterm contribution in 
Eq.~(\ref{eq:vpe1}) subtracts exactly the Born approximation $\delta^{(1)}$ from the 
phase shift \cite{Graham:2009zz}. Again there is subtlety in the presence 
of secondary vacua as there is a direct contribution from the pseudo-potential
as well as from the step function potential inherited from the different 
masses
\begin{equation}
\delta^{(1)}(k)=-\frac{1}{2k}\int_{-\infty}^\infty dx\, V_p(x)\Big|_{x_m}
+\frac{x_m}{2k}\left(m_R^2-m_L^2\right)
=-\frac{1}{2k}\int_{-\infty}^\infty dx\, V_p(x)\Big|_{0}\,.
\label{eq:Born}
\end{equation}
Here the subscript gives the position of the step in the pseudo-potential, $V_p(x)$. 
We stress that this Born approximation for the phase shift is obtained from the full 
fluctuation potential, $V(x)-m_L^2$. In Eq.~(\ref{eq:pseudopot}) we used the unique 
feature that the Born approximation is linear in the fluctuation potential to obtain 
a well-defined integral representation. Especially we have thus found that a particular 
matching point must be chosen and that there is no variation of the Born approximation 
with $x_m$. Of course, for numerical evaluations we always verify that this is the case 
for the full calculation. In total, the vacuum polarization is computed as 
\begin{equation}
E_{{\rm vac}} = \frac{1}{2} \sum_{j}\left(E_{j}-m_L\right) -
\frac{1}{2\pi} \int_{0}^{\infty} {\rm d}k\, \frac{k}{\sqrt{k^2+m_{L}^{2}}}
\left(\delta(k)-\delta^{(1)}(k)\right)\,.
\label{eq:evac}
\end{equation}
In the introduction we mentioned that the renormalization of the VPE in the $\phi^8$ model 
would only be possible for the no-tadpole condition with respect to the full potential. This 
is exactly what Eq.~(\ref{eq:evac}) implies. There is no left-over first order contribution 
to the VPE. Any first order finite renormalization would contain an integral of the 
full fluctuation potential. It cannot be the pseudo-potential since the model by itself
has no information on the matching point. However, the integral over the full fluctuation 
potential does not exist when $m_R\ne m_L$.

For symmetric background potentials $V(-x)=V(x)$ (which implies $m_R=m_L$ and no secondary vacuum) 
a simpler formalism to compute the VPE exists. It makes ample use of analytic properties of 
scattering data and results in~\cite{Graham:2009zz}
\begin{equation}
E_{{\rm vac}}^{S}=\int_{m_{L}}^{\infty} \frac{{\rm d}t}{2\pi} 
\frac{t}{\sqrt{t^2-m_{L}^{2}}} \Bigg[\ln\Bigg\{ g(0,t) \Bigg( 
g(0,t)-\frac{1}{t}g^{\prime}(0,t)\Bigg) \Bigg\}\Bigg]_{1\,.}
\label{eq:Jost}
\end{equation}
The subscript indicates that the Born approximation has been subtracted.
Here $g(x,t)$ is the Jost solution factor that solves the differential equation
\begin{equation}
g^{\prime\prime}(x,t)=2tg^{\prime}(x,t)+V(x)g(x,t)
\end{equation}
for imaginary momenta $t=\imu k$ with boundary conditions $g(\infty,t)=1$ and
$g^{\prime}(\infty,t)=0$. We will use this formalism to verify our results 
in case $V(x)$ is indeed symmetric. It can also be used to 
consider $V(x_0+x)+V(x_0-x)$~\cite{Weigel:2016zbs}. For sufficiently large 
$x_0$ this is a non-interfering superposition and the resulting VPE is twice 
that of $V(x)$~\cite{Graham:1998kz}. Unfortunately, numerically there are 
restrictions on how large $x_0$ can be taken.

The wave-equation~(\ref{eq:waveequa}) is also used to determine the bound state 
energies $E_j<m_L$. We integrate this equation with the initial conditions
$$
\eta_L\,\longrightarrow\,1 \quad{\rm and}\quad 
\eta_L^\prime\,\longrightarrow\, \sqrt{m_L^2-E^2}
\quad {\rm for}\quad x\,\longrightarrow\,-\infty
$$
as well as
$$
\eta_R\,\longrightarrow\,1 \quad{\rm and}\quad 
\eta_R^\prime\,\longrightarrow\, -\sqrt{m_R^2-m_L^2-E^2} 
\quad {\rm for}\quad x\,\longrightarrow\,\infty
$$
from either side. Whenever we tune the energy to $E=E_j$ such that the Wronskian 
$\eta_L\eta_R^\prime-\eta_R\eta_L^\prime$ is zero at any intermediate coordinate 
(preferable $x_m$) we have identified a bound state energy.

\section{The $\mathbf{\phi^8}$ Solitons} 
\label{sec:solitons}

\begin{figure}
\centerline{
\epsfig{file=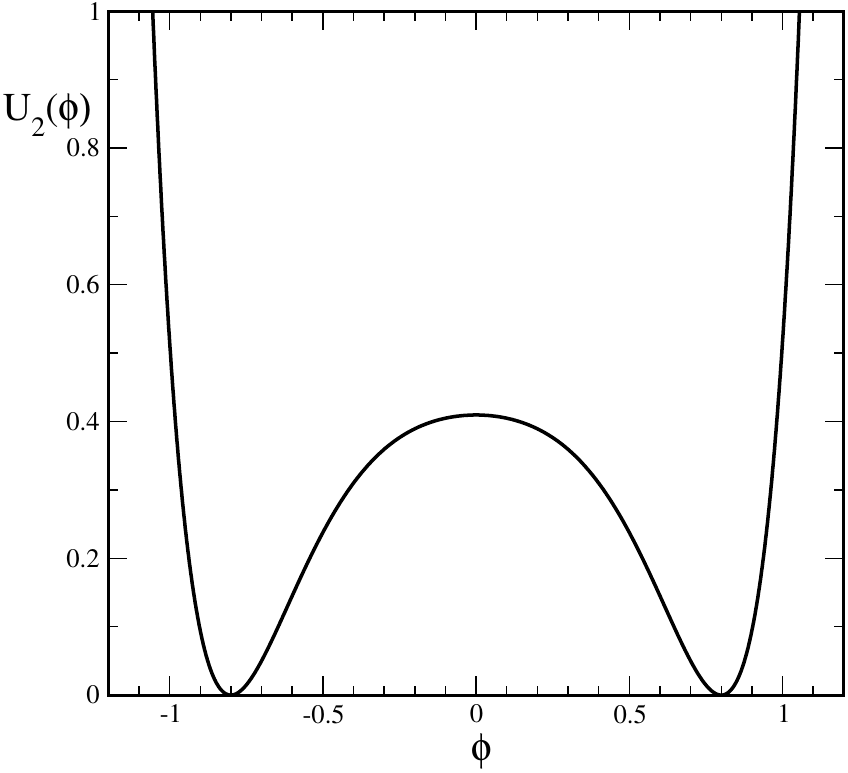,width=5cm,height=3.5cm}~~
\epsfig{file=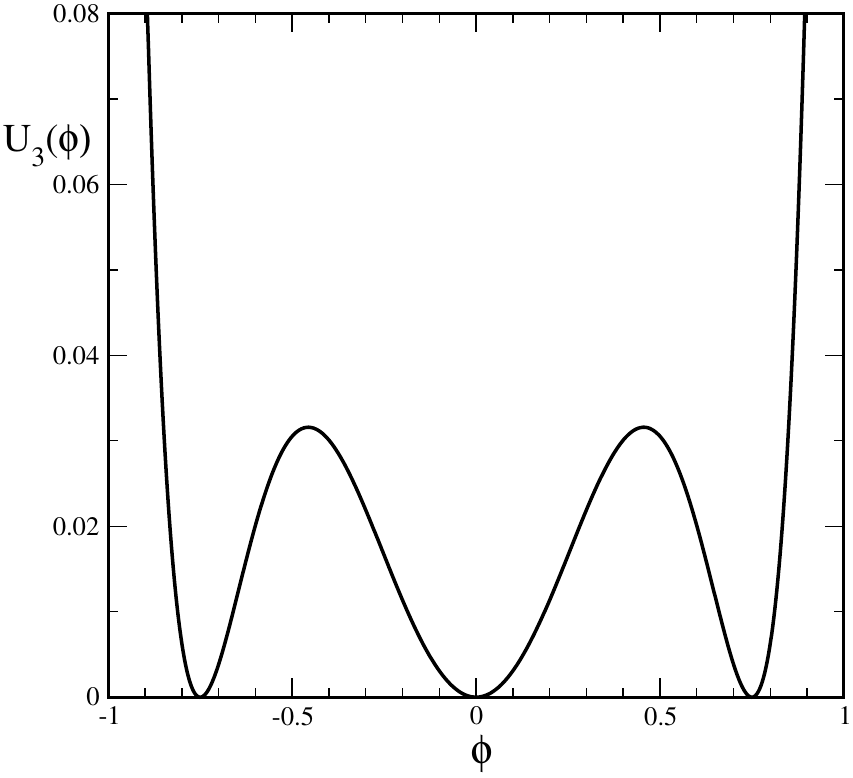,width=5cm,height=3.5cm}~~
\epsfig{file=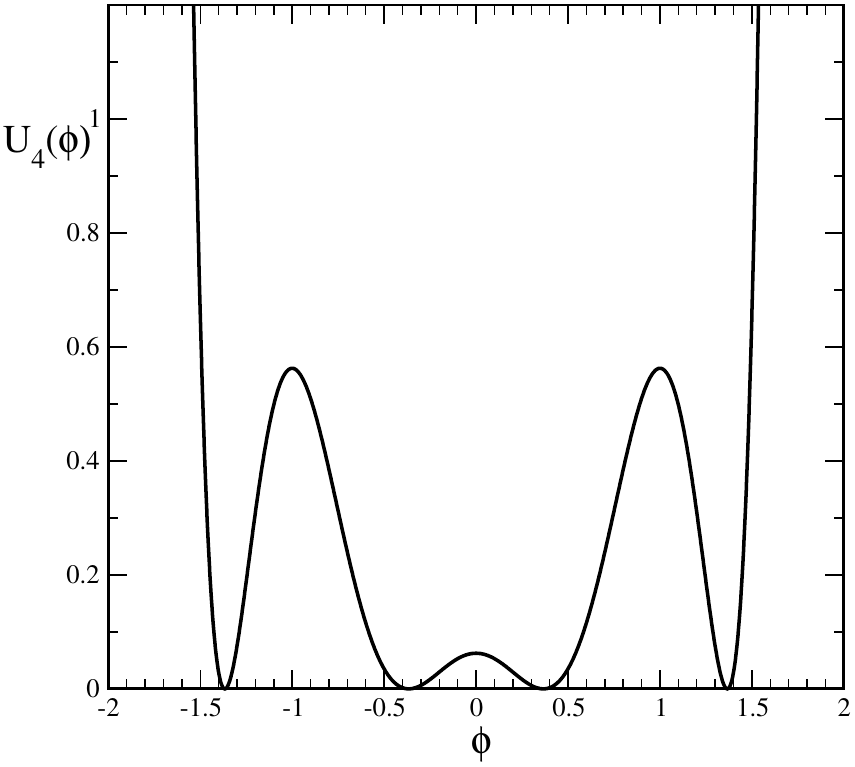,width=5cm,height=3.5cm}}
\caption{\label{fig:Ufield}Field potentials in the $\phi^8$ model: From left to
right for two, three and four degenerate vacua.}
\end{figure}

In this section we discuss the classical solutions from the field potentials in Eq.~(\ref{eq:field1}) 
and list the formulas for the resulting background potentials for the 
fluctuations. To a major part this discussion is a brief review of the findings
from Ref.\cite{Gani:2015cda} with a particular emphasis on the vacuum structures 
and the implications for the fluctuations $\eta(t,x)$. Exemplary graphs for the 
field potentials in the three different sectors are shown in Fig.~\ref{fig:Ufield}.

\subsection{Two degenerate minima: $U_2$}

For the case of $U=U_2$ the kink solution interpolating between $-a$ and $a$ 
is given implicitly by 
\begin{equation}
\pm m_R(x-x_0) = \frac{2a}{b} \arctan \left(\frac{\phi_K}{b}\right)
+ \ln\left(\frac{a+\phi_K}{a-\phi_K}\right)\,,
\label{eq:impsol2deg}
\end{equation}
where $m_R=\sqrt{8}\lambda a(a^2+b^2)$. The corresponding 
classical kink mass is 
\begin{equation}
M_{(-a,a)} = \frac{4\sqrt{2}}{15} \lambda a^3 (a^2 +5b^2).
\end{equation}
The background potential for the fluctuations is symmetric under the spatial 
reflection $x\rightarrow -x$
\begin{equation}
V(x)=V_2(x)=\lambda^2\left\{ 56\phi_{K}^{6} + 60(b^2 -2a^2)\phi_{K}^{4} 
+12(a^4 -4a^2b^2+b^4)\phi_{K}^{2}+4a^4b^2-4a^2b^4\right\}.
\label{eq:backpot2deg}
\end{equation}
An example is shown in the top left panel of figure \ref{fig:backpot}.
By construction $U_2$ only has a primary vacua yielding $m_L=m_R$. 

\subsection{Three degenerate minima: $U_3$}

When $U=U_3$ there are solitons that interpolate between the degenerate 
vacua at $\phi=a$ and $\phi=0$. The implicit solution to the first 
order equation~(\ref{eq:genkink4deg}) with the boundary conditions 
$\phi\,\to\,0$ as $x\,\to\,-\infty$ and $\phi\,\to\,a$ as $x\,\to\,\infty$ is
\begin{equation}
e^{m_R(x-x_0)}=\left(\frac{\sqrt{b^2+\phi_K^2}-b}{\sqrt{b^2 + \phi_K^2}+b}\right)^{\frac{\sqrt{b^2+a^2}}{b}} 
\left(\frac{\sqrt{b^2+a^2}+\sqrt{b^2+\phi_K^2}}{\sqrt{b^2+a^2}-\sqrt{b^2+\phi_K^2}}\right)\,,
\label{eq:impsol3deg}
\end{equation}
with $m_R=2\sqrt{2}\lambda a^2\sqrt{a^2+b^2}$. The resulting classical mass is
\begin{equation}
M_{(0,a)} = \frac{\sqrt{2}}{15} \lambda \left(2(b^2+a^2)^\frac{5}{2}
-b^3(2b^2+5a^2)\right)\,.
\end{equation}
The background potential
\begin{equation}
V(x)=V_2(x)=\lambda^2\Big\{ 56\phi_{K}^{6} + 30(b^2 -2a^2)\phi_{K}^{4} 
+12(a^4 -2a^2b^2)\phi_{K}^{2}+2a^4b^2\Big\}
\label{eq:backpot3deg}
\end{equation}
is not symmetric under the reflection $x\,\leftrightarrow\,-x$, {\it cf.} top right entry of
figure \ref{fig:backpot}. Consequently $V(-\infty)\neq V(\infty)$ and 
$m_L=\sqrt{2}\lambda a^2 b < m_R$. The field potential $U_3$ also has topologically equivalent 
solitons that interpolate between $\phi=-a$ and $\phi=0$. We will not consider them here as 
they are not subject to our choice $m_L\le m_R$. Of course, the numerical results coincide with 
those for the chosen soliton.

We note an interesting relation between the $U_2$ and $U_3$ models. In the limit $b\,\to\,0$ the 
soliton of the $U_2$ model separates into two structures. In one region the soliton has a kink
shape that connects $\phi=-a$ and $\phi\approx0$. In the second region the soliton then
connects $\phi\approx0$ and $\phi=+a$. The smaller $b$, the further apart are these regions. 
Each of the two structures then is similar to a soliton of the $U_3$ model. Stated otherwise, any 
peculiar feature of the $U_3$ model (like instabilities) should also be seen in the $U_2$ model
when the limit $b\,\to\,0$ is considered.

\subsection{Four degenerate minima: $U_4$}

For the model with four degenerate minima, $U_4$ in Eq.~(\ref{eq:field1}) we take $b>a>0$. 
The four degenerate minima are at $\phi(x,t)= \pm a $ and $\phi(x,t)=\pm b$. The interesting 
feature is that there are two distinct soliton solutions with inequivalent topology. 

\subsubsection{Kink interpolating between $-a$ and $+a$}

In this sector, the field is constrained to be $\vert \phi \vert < a$. The implicit solution
for the kink is
\begin{equation}
e^{\pm m_L(x-x_0)}=\left(\frac{a+\phi_K}{a-\phi_K}\right)\left(\frac{b-\phi_K}{b+\phi_K}\right)^{\frac{a}{b}} \,,
\label{eq:impsol4degaa}
\end{equation}
where $m_L = \sqrt{8} \lambda a(b^2-a^2)$. It has the classical mass 
\begin{equation}
M_{(-a,a)}=\frac{4\sqrt{2}}{15} \lambda a^{3}\left(5b^{2}-a^{2}\right)\,.
\end{equation}
The background potential for the fluctuations
\begin{equation}
V(x)=V_4(x)=\lambda^2\Big\{ 56\phi_{K}^{6} - 60(a^2 + b^2)\phi_{K}^{4} 
+12(a^4 +4a^2b^2+b^4)\phi_{K}^{2}-4a^4b^2-4a^2b^4\Big\}
\label{eq:backpot4degaa}
\end{equation}
is symmetric under the spatial reflection $x\,\leftrightarrow\, -x$. This, of course, implies 
$m_R=m_L$. A typical example is shown in the bottom left panel of figure \ref{fig:backpot}.

\subsubsection{Kink interpolating between $a$ and $b$ (or $-b$ and $-a$)}

In this case the field is constrained by $a<\vert \phi \vert < b$ and the 
implicit solutions reads
\begin{equation}
e^{\pm m_L(x-x_0)} = \left(\frac{\phi_K-a}{\phi_K+a}\Bigg)
\Bigg(\frac{b+\phi_K}{b-\phi_K}\right)^{\frac{a}{b}} \label{eq:impsol4degab},
\end{equation}
with $m_L$ as above. For the kink in this sector the classical mass of the kink is
\begin{equation}
M_{(a,b)}=\frac{\sqrt{8}}{15} \lambda (b-a)^3\left(a^2+3ab+b^2\right).
\end{equation}
Formally and by construction, the background potential for the fluctuations 
is that of Eq.~(\ref{eq:backpot4degaa}). However, since the kink is not 
invariant under spatial reflection, the background potential is not either
and we obtain different mass parameters: $m_R=bm_L/a>m_L$.

\section{Numerical Results}
\label{sec:results}

In this section we present the numerical results for the VPEs for the different topological
sectors discussed above. As mention earlier, we take the model parameter $\lambda=1$. This 
is legitimate as long as we are only interested in the VPE. In general, $\lambda$ serves as 
a loop-counting parameter and the classical mass scales inversely with $\lambda$ while the 
VPE is only proportional to $m$, where $m$ is a mass parameter\footnote{For this to be
correct, the dimensionless parameters $a$ and $b$ must be written as $a=\alpha\sqrt[3]{m/\lambda}$
and $b=\beta\sqrt[3]{m/\lambda}$ where $\alpha$ and $\beta$ do not vary with $m$ or $\lambda$. 
With this scaling the quadratic mass type term in $U(\phi)$ does not contain the coupling constant
$\lambda$. In turn the kink profile, $\phi_K$ also scales like $\sqrt[3]{m/\lambda}$ and the 
prefactor $\lambda^2$ cancels in $V(x)$, {\it cf.} Eq.~(\ref{eq:backpot4degaa}). Then the
classical mass scales as $\left(m/\lambda\right)^{2/3}$.} in the potential $U(\phi)$.

We first note that for the classical soliton and the discrete bound state energies we have 
reproduced all numerical results reported in Ref.\cite{Gani:2015cda}. Of course, scattering 
solutions were not considered in that paper. When computing the phase shifts we have furthermore 
verified that the number of bound states $n_{\rm bs}$ agrees with Levinson's theorem in one 
space dimension~\cite{Barton:1984py,Graham:2001iv} according to which the phase shift at 
zero momentum is $\delta(0)=\pi(n_{\rm bs}-1/2)$. Numerically there actually is a non-trivial
ingredient as the phase shift computed from Eq.~(\ref{eq:phase}) is always between $-\pi/2$ 
and $+\pi/2$ with jumps of $\pi$ whenever the determinant of the scattering matrix passes 
negative one. These jumps are eliminated by adding appropriate multiples of $\pi$ to the phase
shift with the boundary condition $\delta(\infty)=0$. In this way we not only get a continuous
phase shift, but also agreement with Levinson's theorem in all our simulations.

The computation of scattering data is hampered by the fact that the soliton profiles are only 
available in form of implicit expressions, {\it cf.} Eq.~(\ref{eq:impsol2deg}). Numerically 
we solve them by the method of nested intervals. Though that procedure is sufficiently 
efficient for quantities on the classical level, the differential equations of the scattering
problem, Eqs.~(\ref{eq:modwvequa}) and~(\ref{eq:parametequa}), are treated within an adaptive 
step size algorithm and {\it a priori} it is unknown at which coordinate the profile functions
are needed. Therefore the nested intervals procedure must be applied at every coordinate 
requested by the algorithm. This is numerically time consuming.

\begin{figure}
\centerline{
\epsfig{file=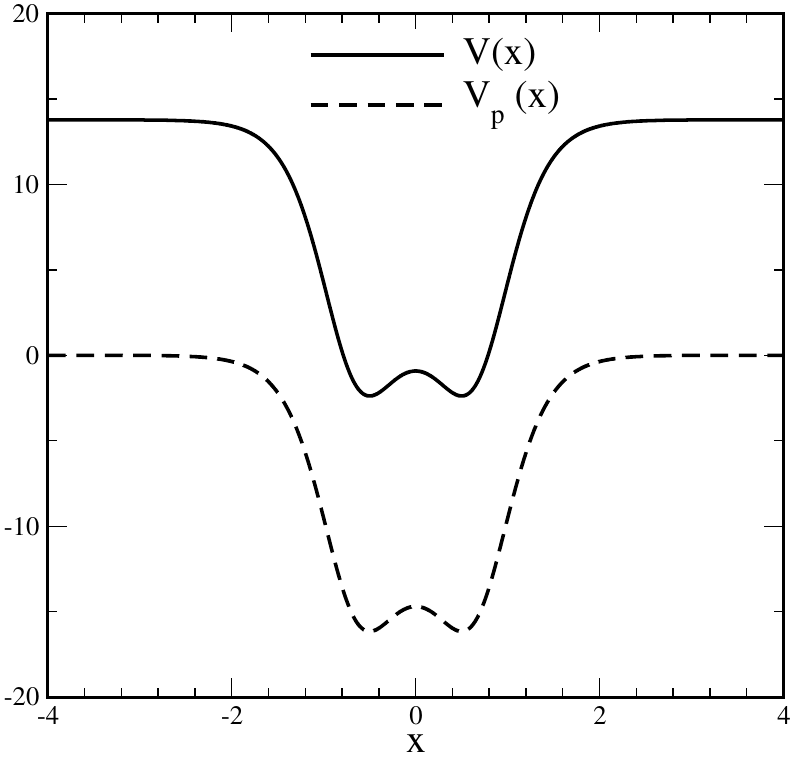,width=7cm,height=3.5cm}\hspace{1cm}
\epsfig{file=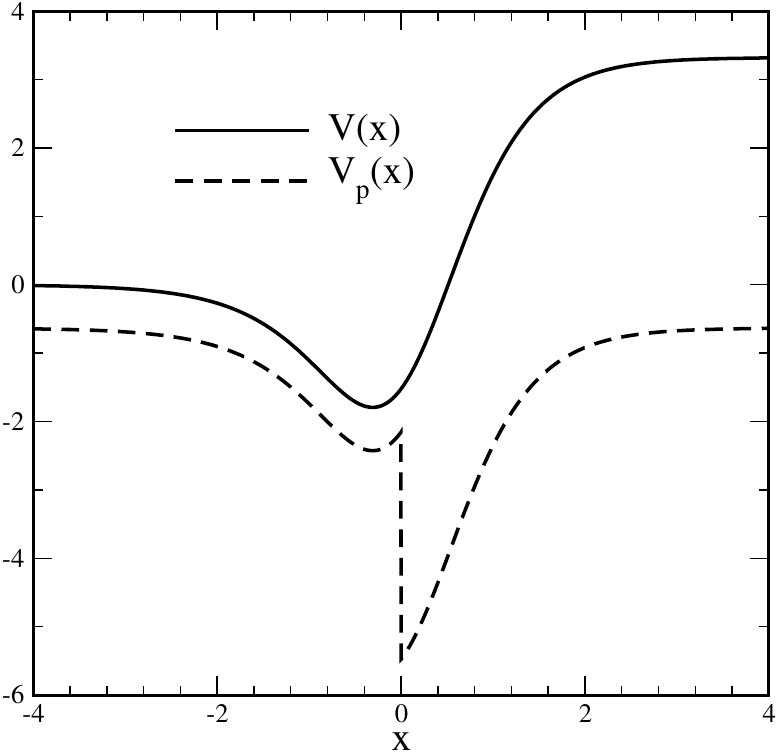,width=7cm,height=3.5cm}}
\centerline{
\epsfig{file=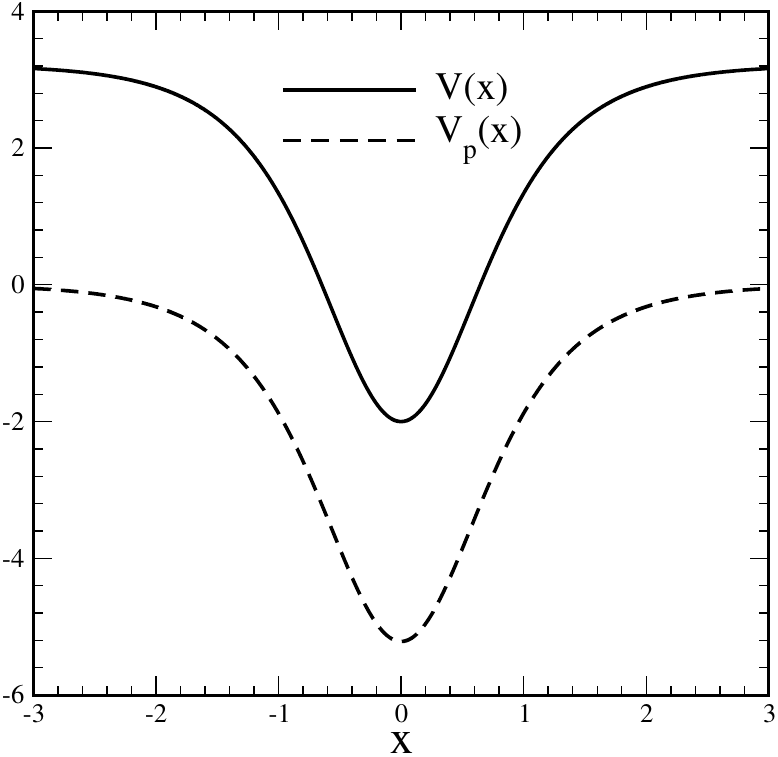,width=7cm,height=3.5cm}\hspace{1cm}
\epsfig{file=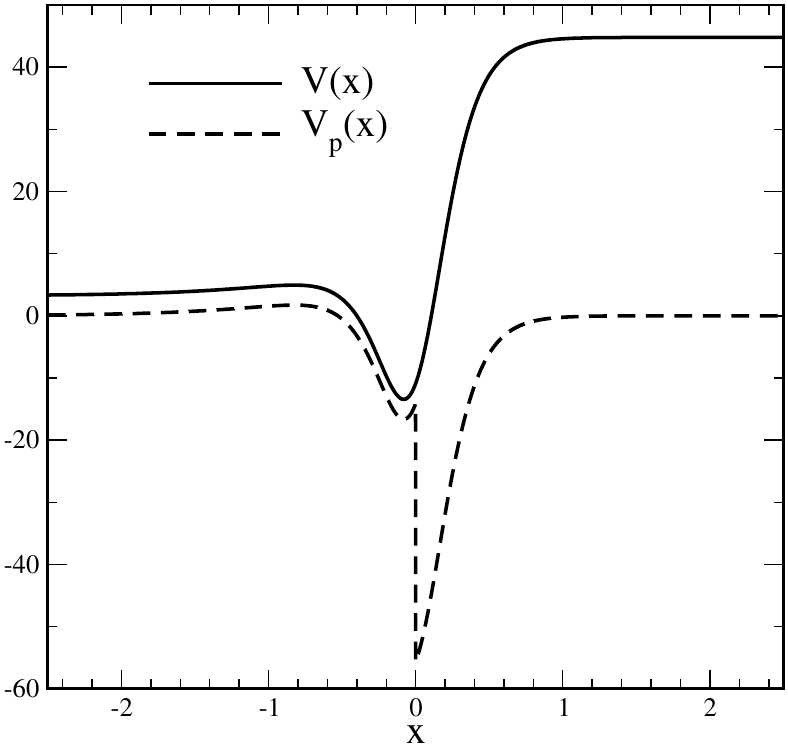,width=7cm,height=3.5cm}}
\caption{\label{fig:backpot}Background- and pseudo-potentials from solitons in the various sectors. 
All cases have $x_0=0$. Top row: models with two (left panel) and three (right panel) degenerate minima. 
Bottom row: model with four degenerate minima and symmetric (left panel) and non-symmetric
(right panel) solitons.}
\end{figure}

In figure \ref{fig:backpot} we show examples for the background potential that enter
the differential equations for the scattering data.

\subsection{Two degenerate minima: $U_2$}

The model with two degenerate minima of the field potential is conceptually that of the 
standard $\phi^4$ kink model as the two vacua have equal curvature. Hence, as a proof of
concept, we only compute the VPE for a single set of parameters that we take as in 
Ref.~\cite{Gani:2015cda}: $a=\frac{4}{5}$ and $b=1$. This translates into $m_{L}=m_{R}=3.710$.
The major interesting aspect is that there four bound states. Accordingly the phase shift at 
zero momentum approaches $7\pi/2$ as seen in the top left entry of figure \ref{fig:phase}.
\begin{table}
\centerline{
\begin{tabular}{c|c|c|c||c|c|c}
\multicolumn{4}{c||}{bound state energies} 
& $E_{\rm bind.}$ & $E_{\rm scat.}$ & $E_{\rm vac.}$\cr
\hline
0.0 & 2.067 & 3.192 & 3.689 & -2.947 & 1.3722 & -1.575 \cr
\end{tabular}}
\caption{\label{t4}Numerically obtained energies from scattering solutions in
the model with two degenerate minima and the
parameters $a=\frac{4}{5}$ and $b=1$. Note that the threshold is at $m_{L}=m_{R}=3.710$.
The entries $E_{\rm bind.}$ and $E_{\rm scat.}$ denote the bound state and continuum
contributions to the VPE, {\it i.e.} the two distinct terms in Eq.~(\ref{eq:evac}).}
\end{table}
The bound state energies as well as the various contributions to the VPE and the VPE itself are 
listed in table \ref{t4}. As for the standard kink, the leading quantum correction is negative.
We have confirmed the VPE result using the Jost function formalism of Eq.~(\ref{eq:Jost}). The
numerical result from that calculation is $E^S_{\rm vac.}=-1.574$.

\subsection{Three degenerate minima: $U_3$}

Again we adopt the model parameters of Ref.~\cite{Gani:2015cda}: $a=\frac{3}{4}$ and $b=1$.
Even though the model produces the translational zero mode as the only bound state, this 
model is nevertheless more interesting than the one with two degenerate minima. The reason is that with 
the additional minimum we now have primary and secondary vacua. This is most obvious from the 
meson masses (curvatures): $m_L=0.7955$ and $m_R=1.9887$. This yields $-0.398$ for the bound 
state contribution to the VPE irrespective of $x_0$. Furthermore, the phase shift exhibits the 
typical threshold cusp at $k=\sqrt{m_R^2-m_L^2}$ as seen on the top right in figure \ref{fig:phase}. 
Obviously the phase shift at zero momentum, $\delta(0)\approx\pi/2$ is consistent with 
Levinson's theorem.

Most importantly the appearance of the secondary vacuum at spatial infinity induces a 
translational variance of the VPE as seen in table \ref{t3}. As the kink is shifted towards the
primary vacuum, the region with the secondary vacuum increases and low-lying modes disappear.
Consequently the VPE decreases.
\begin{table} 
\caption{\label{t3} The VPE as a function of the center of the kink $x_0$ in the model with 
three degenerate minima.}
\begin{center}
\begin{tabular}{c| c c c c c c c c c }
 $x_{0}$ & $1.00$  & $0.75$  & $0.50$  & $0.25$  
         & $0$      & $-0.25$  & $-0.50$  & $-0.75$ & $-1.00$   \\ \hline 
$E_{\rm vac}$   & $0.254$  & $0.215$  & $0.176$  & $0.138$  
         & $0.0986$ & $0.0595$ & $0.0211$ & $-0.0179$ & $-0.0568$   \\
\end{tabular}
\end{center}
\end{table}
As matter of fact, there is no bound to this decrease and the VPE can take any arbitrarily large
negative value. Hence for any fixed loop-counting parameter $\lambda$, there will be an $x_0$ 
such that the total energy is negative and the quantum corrections destabilize the soliton.

\subsection{Four degenerate minima: $U_4$}

For the numerical exploration we again adopt the relevant parameters from Ref.\cite{Gani:2015cda}: 
$a=(\sqrt{3}-1)/2$ and $b=a+1$. This translates into $m_L=1.793$ and $m_R=6.692$.

\begin{figure}
\centerline{
\epsfig{file=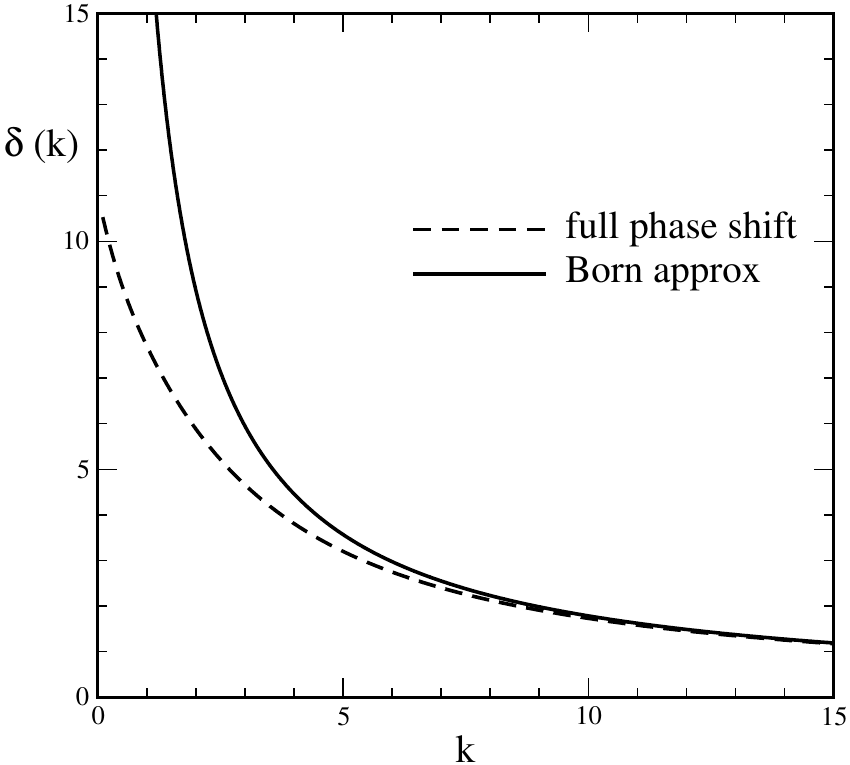,width=7cm,height=3.5cm}\hspace{1cm}
\epsfig{file=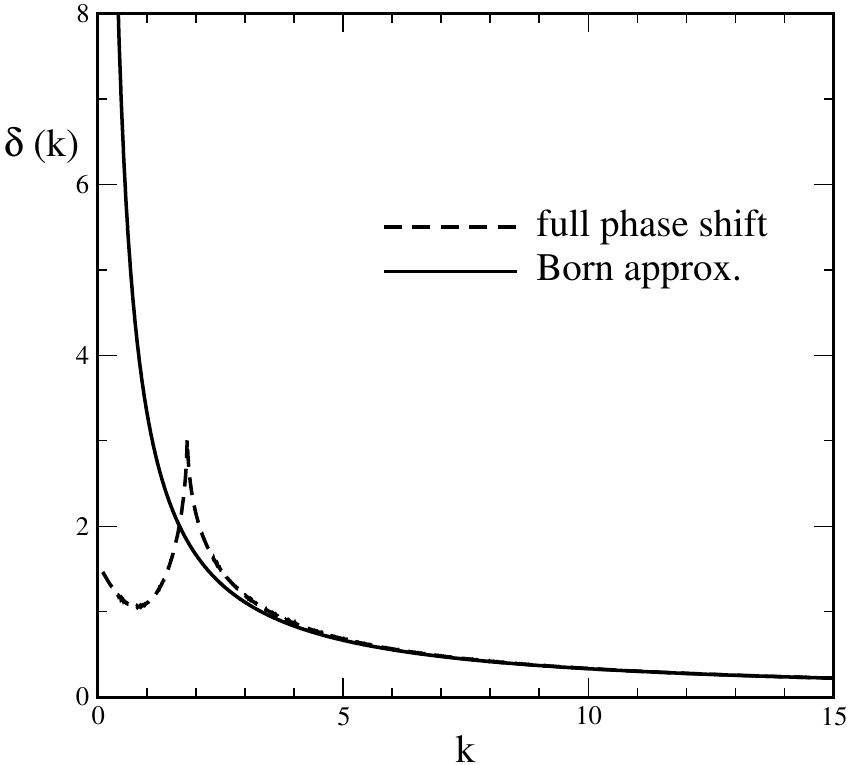,width=7cm,height=3.5cm}}
\centerline{
\epsfig{file=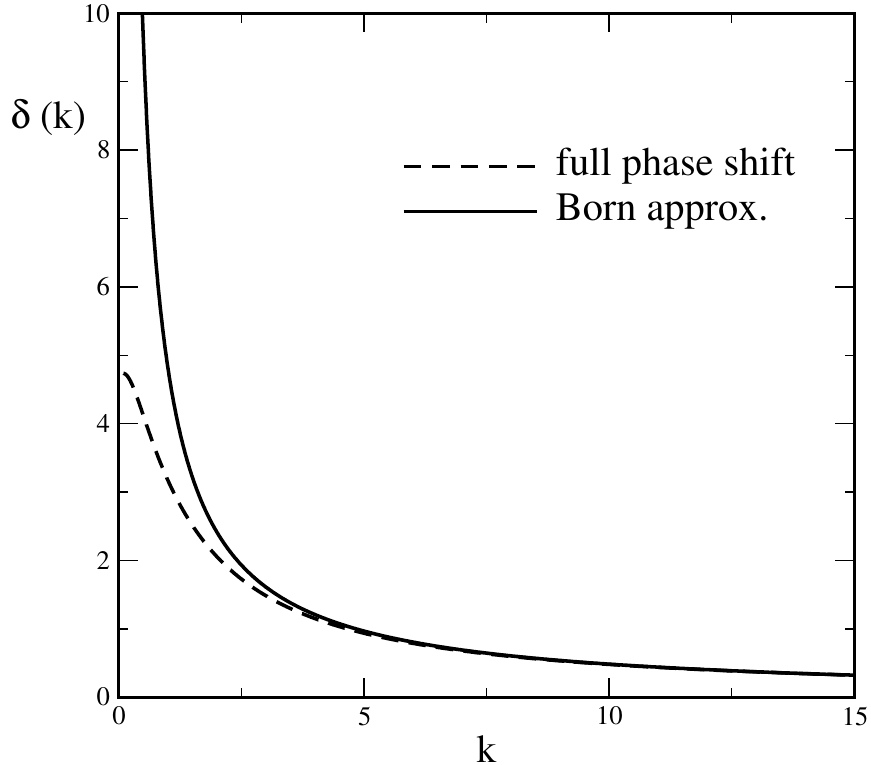,width=7cm,height=3.5cm}\hspace{1cm}
\epsfig{file=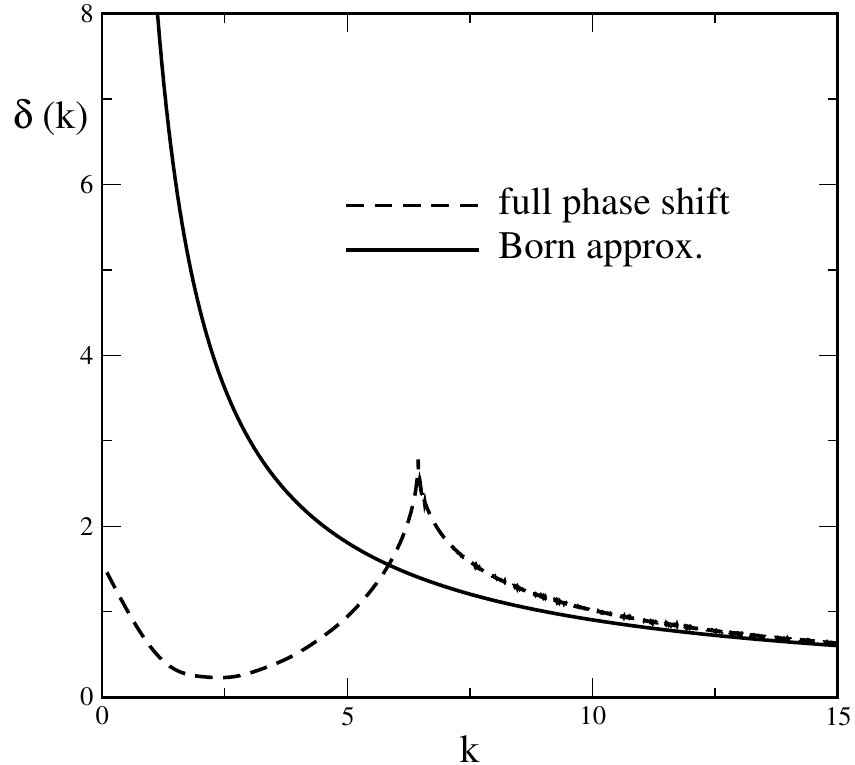,width=7cm,height=3.5cm}}
\caption{\label{fig:phase}Sample phase shifts in the various version of the $\phi^8$ model. Top row: two (left) and
three (right) degenerated vacua. Bottom row: four degenerate vacua for the sectors $(a,a)$ and $(a,b)$ in the left
and right panels, respectively. The Born approximations are computed from Eq.~(\ref{eq:Born}).}
\end{figure}

In the topological sector $(-a,a)$ the background potential is symmetric under the spatial
reflection $x \rightarrow -x$ as seen in the bottom left entry of figure~\ref{fig:backpot}.
In this regime the squared masses for the vacua are equal and consequently $q=k$ in 
Eq.~(\ref{eq:parametequa}). Numerically we confirm that the determinant of the scattering 
matrix is invariant under the translation $x\rightarrow x-x_0$ and that the VPE is independent 
of $x_0$. We present the bound state energies and the corresponding VPE for this kink solution 
in table~\ref{t1}.
\begin{table}
\begin{center}
\begin{tabular}{c|c|| c| c| c}
\multicolumn{2}{c||}{bound state energies}
& $E_{\rm bind.}$ & $E_{\rm scat.}$ & $E_{\rm vac.}$\cr
\hline
~~~$0.0$~~~~~& $1.644$  &  $-0.971$  & $0.389$ & $-0.582 (0.583)$ 
\end{tabular}
\end{center}
\caption{\label{t1}The bound states energies of the kink of Eq.~(\ref{eq:impsol4degaa})
and its VPE. The last entry in parenthesis is computed 
via $E_{\rm vac}^{S}$ in Eq.~(\ref{eq:Jost}).}
\end{table}
Besides the mandatory translational zero mode the soliton has a shape mode slightly below 
threshold. We show the sum of the eigenphase shifts entering this computation in the bottom 
left panel of figure~\ref{fig:phase} and read off $\delta(0)=\frac{3\pi}{2}$ confirming the 
existence of two bound states via Levinson's theorem.

In the case of the topological sector $(a,b)$ we only have one bound state, the translational zero 
mode. Consequently $\delta(0)=\frac{\pi}{2}$ as observed in bottom right entry of figure~\ref{fig:phase}.
The binding energy is $\sum_{j}(E_{j}-m_L)=-0.896$. 

In this topological sector we have $m_{R}(6.692)\neq m_{L}(1.793)$ and the existence of a 
threshold is also reflected by the cusp in the phase shift. More importantly, the different masses
correspond to primary and secondary vacua and, as seen from table~\ref{t2} the VPE depends 
in the position of the soliton.

\begin{table} 
\begin{center}
\begin{tabular}{c| c  c  c c c }
$x_{0}$     &  $0.50$   & $0.25$  &  $0$      &  $-0.25$  & $-0.50$   \\ \hline 
$E_{\rm vac}$     &  $2.628$   & $1.869$  &  $1.111$  &  $0.348$ &  $-0.407$ \\
\end{tabular}
\end{center}
\caption{\label{t2} The VPEs as a function of the center of the kink $x_0$ of the topological 
sector $(a,b)$ in the four degenerate minima regime.}
\end{table}
Obviously there is no lower bound to the VPE and the emergence of a secondary vacuum again causes quantum
effects to destabilize the soliton.

\subsection{Moderate differences in the curvatures}

The $\phi^8$ model would be a perfect sample to study the dependence of the VPE on the 
topological charge because the model with four minima admits soliton solutions with different
charges without changing any model parameter. These are kinks that connect $(a,b)$ and $(-a,a)$ 
respectively. One would simply compare the calculated VPEs. Unfortunately, the $(a,b)$ 
kink induces a translational variance of the VPE and no unique value can be assigned.
To nevertheless get a rough idea, we consider parameters $a$ and $b$ leading to $m_L=2.0$ 
and $m_R=2.5$ such that the difference in the curvatures is small to moderate. The 
topological charges scale as $Q_{(-a,a)}/Q_{(a,b)}=8$ and the classical masses
approximately follow that behavior $M_{(-a,a)}/M_{(a,b)}=8.44$. Yet the $(-a,a)$ soliton
is not stable energetically but topologically since $b>a$.

The VPE for this $(-a,a)$ soliton is $-1.242$. The translational variance of the VPE for the
$(a,b)$ soliton is listed in table \ref{t5}.
\begin{table}
\centerline{
\begin{tabular}{c| c c c c c}
$x_0$ &  -1.0  & 0.25  &  0.0  & 0.25  &  1.0 \cr
\hline
$E_{\rm vac}$ & -0.578 & -0.553 & -0.550 & -0.542 & -0.528
\end{tabular}}
\caption{\label{t5}Translational variance of VPE for the $(a,b)$ soliton for 
$m_L=2$ and $m_R=2.5$ in the model with four minima.}
\end{table}
As expected there is only a mild (linear) dependence of the VPE on $x_0$ that we 
fit to $E_{\rm vac}\approx \overline{E}_{\rm vac}+\epsilon_{\rm vac}x_0$ with
$\overline{E}_{\rm vac}=-0.550$ and $\epsilon_{\rm vac}=0.025$. Assuming 
$\overline{E}_{\rm vac}$ as a reasonable measure for the VPE of the $(a,b)$ soliton
we see that the VPEs in different topological sectors do not scale with the
topological charge. Since this is different from the classical masses, the VPE
can have significant effects when estimating the binding energies of solitons 
with large topological charges. To compare the VPE of the decaying soliton with
that of the decay products we need to compare $8\overline{E}_{\rm vac}\approx-4.40$
with the VPE of the $(-a,a)$ soliton and conclude that including the VPE has the
potential to reduce the binding energy significantly. A scenario that was 
also seen for the $H$-dibaryon in the Skyrme model~\cite{Scholtz:1993jg}.
In the present model the total energies depend on the particular value of
the loop-counting parameter $\lambda$ for which we, unfortunately, have no
empirical input. Nevertheless we think that our results may alter 
the picture obtained for binding energies of nuclei in phenomenological 
soliton models~\cite{Feist:2012ps} once the VPE is included which is hampered
by those models not being renormalizable.

\subsection{Transition between the two and three degenerate minima models}

The soliton in the model with three minima can be viewed as a limiting case of the one in 
the model with only two minima. As we decrease $b$ in the $U_2$ model, the $(-a,a)$ configuration
disintegrates into two separated $(-a,0)$ and $(0,a)$ structures each being similar to the
kinks in the model with three minima. Changing that separation translates into changing the 
center of the $(0,a)$ kink in the model with three minima. We hence expect that the VPE of 
the model with only two minima becomes large and negative as we tune $b$ towards zero. This 
is exactly the behavior of the data listed in table \ref{t6}, even though there is a small 
increase of the VPE as we decrease $b$ in the moderate regime.
\begin{table}
\centerline{
\begin{tabular}{c| c c c c c c c c c}
$b$ &  1.0  & 0.7  &  0.5  & 0.2  &  0.1  & 0.08  & 0.07 & 0.06 & 0.05\cr
\hline
$E_{\rm vac}$ & -1.574 & -1.307 & -1.277 & -1.868 & -3.162 
& -3.846 & -4.341 & -5.006 & -5.947
\end{tabular}}
\caption{\label{t6}VPE as function of $b$ in model with two minima.}
\end{table}
The $b\to0$ limit of the $U_2$ potential takes the shape of a wide well. On the other
hand, two widely separated $(-a,0)$ and $(0,a)$ structures form a potential barrier
in the $U_3$ model. The height of this barrier equals the depth of the well. With the 
no-tadpole renormalization scheme the second
order contribution from the potential dominates the VPE. This second order piece is 
the same for the well and the barrier. Hence we can indeed expect the two scenarios
in the $U_2$ and $U_3$ models to yield similar results for that limiting scenario.

\section{Conclusion}
\label{sec:conclude}

In this project we studied the leading (one-loop) quantum corrections to classical
energies of solitons in the $\phi^8$ model in one space and one time dimensions. We utilized
spectral methods for this study. These methods merely require to compute the scattering
data for fluctuations about the solitons. No other approximation or truncation is needed. 
The Born approximation enters the calculation only as a technical tool when renormalizing 
the ultraviolet divergences.

Central to our study has been the question of whether or not these corrections destabilize 
classically stable solitons. The key issue to this problem is the existence of secondary
vacua. Here the $\phi^8$ model is unique as it possesses these vacua, but not all soliton
solutions approach these vacua at spatial infinity. Our simulations confirm the earlier
conjectured picture: Whenever the soliton can connect to a secondary vacuum, increasing
the coverage of that vacuum (as a portion in space) reduces the quantum energy without
lower bound. This signals soliton destabilization at the quantum level. This scenario is 
furthermore supported by computing the quantum correction in a model that only has a primary 
vacuum but a certain limiting choice of the model parameters approaches a model that has a 
secondary vacuum. The quantum correction to the energy of the soliton is also found to diverge 
in that limit. We stress, however, that destabilization cannot be solely attributed to the field 
potential but also requires a soliton configuration that connects primary and secondary vacua.

We have also obtained indications that the (one-loop) quantum correction to the energy does not 
scale with the topological charge, even though translational variance in the presence of 
secondary vacua prohibits robust conclusions. In the particle picture of soliton models 
the topological charge is identified as the particle number and the classical energy, that
approximately scales like the charge, as the particle mass. This predestines these models to
predict binding energies of compound objects, such as nuclei. Our results hence suggest that 
quantum corrections should not be ignored when estimating such binding energies.

\acknowledgments
H.\@ W.\@ is supported in part by the National Research Foundation of
South Africa (NRF) by grant~109497.

%\acknowledgments

\end{document}